\documentclass[onecolumn]{emulateapj}

\def\kms{~km~s$^{-1}$\ }

\def\arcs{\char'175\ }
\def\arcsec{\char'175 }

\def\hub{\ifmmode H_\circ\else H$_\circ$\fi}

\shorttitle{Resolved Stars in Virgo}
\shortauthors{Caldwell }

\begin{document}

\slugcomment{to appear in ApJ}

\title{Color-Magnitude Diagrams of Resolved Stars in Virgo Cluster Dwarf Galaxies }


\author{Nelson Caldwell}
\affil{Smithsonian Astrophysical Observatory, 60 Garden Street, Cambridge, MA 02138 }
\email{caldwell@cfa.harvard.edu}



\begin{abstract}
The Advanced Camera for Surveys (ACS) on HST has been used to image two
fields in the core of the Virgo cluster 
that contain a number of dwarf elliptical galaxies.  The 
combined F555W and F814W images have resolved red giant stars in these
galaxies, down to 1 mag below the giant branch tip.  Two of the
galaxies were targeted because of their extremely low central surface brightnesses
($\rm{B}_{\rm o} > 27.0$), thus the successful resolution into stars confirms
the existence of such tenuous galaxies.  Red giant stars were also found
that are not ostensibly associated with any galaxy. Color-magnitude diagrams 
in V and I have been derived for the five dwarfs, as well as the 
halo of a nearby spiral galaxy  and the intracluster stars in the two
fields. These diagrams were used to derive distances and metallicities via 
the magnitude of the red giant branch tip, and the mean color of the
giant branch.   The mean abundances of stars in the dwarfs range from 
$-1.2<[\rm{Fe/H]}<-2.4$, and fall along the relation between galaxy luminosity
and metallicity found for Local Group and M81 group dwarf ellipticals.
[Fe/H] does not appear to be well-correlated with 
galaxy surface brightness, as the two extremely low surface brightness galaxies
do not have extreme abundances.  The mean distance modulus of the six Virgo galaxies
is $31.0\pm0.05$, or $16.1\pm0.4$ Mpc, whereas that for the intracluster
stars in those fields is $31.2\pm0.09$  ($17.4\pm0.7$ Mpc).

\end{abstract}




\section{Introduction}
Dwarf galaxies in nearby clusters, particularly the dominant non-star forming galaxies,
have been the subject of numerous
studies, 
with the topics ranging from morphology, distribution and 
kinematics within the clusters, stellar populations, internal kinematics, 
and globular cluster content \citep{1987AJ.....94..251B,2002MNRAS.333..423T,2003AJ....125...66C,
2003AJ....126.1794G,2003ApJ...587..605M,2005A&A...442...85S}.  Such work
on the 
Local Group dwarf ellipticals
and spheroidals (dE and dSph) 
of course began much earlier and is now comprehensive. Moreover, these galaxies 
are near enough that properties
of individual stars can be studied extensively as well.  Resolved star
analyses have recently begun on the dwarf galaxy populations outside the
Local Group in
nearby groups such as the M81, Sculptor, and Cen A groups \citep{2004PASA...21..366D,
2001A&A...371..487J,2001A&A...375..359K,2002A&A...383..125K,
2003A&A...404...93K,2006A&A...448..983R,1998AJ....115..535C}.
The next step would seem to be to resolve stars in the vast population
of dwarfs found in the large clusters such as Virgo, Fornax and Coma,
thus allowing detailed analyses of dwarf galaxy populations similar to
the studies of the nearby dwarfs.  \citet{1998Natur.395...45H} used WFPC2 on
HST to study one bright Virgo dE. This paper presents an attempt using
the Advanced Camera for Surveys (ACS).

A CCD imaging study of several selected areas in
the Virgo cluster has been conducted by \citet{VirgoCCD}, using the
KPNO 4m telescope and the Mosaic CCD camera with B and R filters.
The long exposures used allowed the detection depth in magnitude of dwarf ellipticals
to be pushed about 1 mag fainter than previously done photographically \citep{1987AJ.....94..251B}, and also 1 mag fainter in surface brightness than other recent CCD 
surveys (e.g., \citet{2002MNRAS.333..423T}). Using visual and automatic techniques, 
a new catalog of dwarfs was made, containing many which have physical parameters placing them
in the category of the fainter Local Group dSph galaxies, and some of 
which are unknown in the Local Group, in the sense that their surface brightnesses
are lower than the nearby dwarfs.  Moreover, such galaxies are large in radius, 
and hence their luminosities are actually brighter than the faintest dSph in the
Local Group.

A question remains however, of the reality of the extremely low surface brightness
galaxies.  It is curious that such galaxies can exist in the center of the
Virgo cluster, where galaxy interactions and tides might have a destructive 
effect, depending on the orbits the dwarfs follow within the cluster.
On ground based images, a tight, distant group of galaxies can often appear
similar to a low surface brightness galaxy at 15 Mpc.  Imaging with HST would
obviously settle the matter, and perhaps provide a bonus of giving 
an accurate distance to the galaxies via the magnitudes of the red giant branch stars
(RGB),
if they can be resolved.  A color-magnitude diagram (cmd) would further allow
an estimate of the giant branch color, which is determined by metal-abundance.
The abundance for an extremely low surface brightness galaxy in turn can
be compared with similarly determined abundances for dE and dSph galaxies
in the Local Group and other nearby groups.  A well-defined relation between
the luminosity and the mean abundance has been established for dE and dSph galaxies. 
Does this also hold true for the extreme surface brightness galaxies, which
reside in the densest regions of the Virgo cluster?

\section{Observations}
The advent of ACS on HST offers another
magnitude of depth from the work done with WFPC2 resolving stars and forming
color-magnitude diagrams for old populations in galaxies of
the Local Group,  the M81 group other nearby environs (e.g.,\cite{1996AJ....112.2576D} ). With 
moderate exposure times on WFPC2, depths of I=26 mag were achieved. Virgo
cluster galaxies are thought not to be closer than about 15 Mpc, thus
typical red giant tip stars with absolute magnitude of M$_{\rm I} \approx -4$ could 
appear at apparent I=26.9, plus an extinction of around 0.05 mag. 
Using WFPC2 exposures totaling 32.2 ksec, \citet {1998Natur.395...45H} detected stars down to I$\sim$ 28 in IC3338, 
a relatively bright Virgo dE.
Therefore, it seemed likely that ACS could resolve red giant stars in Virgo early type 
galaxies, with fairly long exposure times. A further item to consider 
is that at
15 Mpc, the surface number density of stars would lead to confusion for
normal high surface brightness galaxies, except in the outermost parts.
The extremely low surface brightness galaxies identified in the CCD survey then 
provide an excellent target for a project to resolve Virgo red giant stars.

\subsection{Selection of Targets} 
From the more than one hundred low surface brightness Virgo dwarfs 
newly cataloged in \citet{VirgoCCD}, two were selected for
ACS imaging, lying in the fields designated ``N'' and ``SW2''.  
The dwarf galaxy SW2 lsb31 was chosen because its surface
brightness is low enough so that it was missed in the \citet{1985AJ.....90.1681B} 
photographic survey, yet not
so low that its existence was in doubt.  That dwarf lies between two
larger galaxies, one an SB spiral (NGC4407) and the other a much 
brighter, nucleated dE (VCC871). While the centers of those two do not
fall on the small ACS field, their extended stellar halos do show up
in the images.  N lsb10 is one of the most extreme cases found in the CCD survey, 
with a central B surface brightness of about 27.3 mag arcsec$^{-2}$.  It is also
large, with a limiting radius (at 30 mag arcsec$^{-2}$) of 30\arcsec, or 2 kpc if it is 
indeed in the Virgo cluster.   The question remained however of its
true nature - while the object was definitely real, showing up on 
multiple exposures and in both colors, the possibility that it was
a background object, as mentioned above, remained.  This field also contained
two other objects: a low luminosity, but
higher surface brightness dE, cataloged in \citet{1985AJ.....90.1681B} (VCC941), 
and a small galaxy that was
not identified by eye in the CCD survey as a dE member, because its appearance
in the ground based images is too similar to distant background 
galaxies.  And while it was cataloged in the automatic survey as a possible 
dE, its appearance on the ACS images as a resolved dSph galaxy in the Virgo
cluster with a total luminosity of V=-9.6 was entirely due to good fortune.
As an indication of the range of structures in dwarf ellipticals, Table \ref{data.tab}
contains the effective radii of the galaxies ($\Re _{\rm eff}$). Note that
while N lsb10 is nearly a magnitude fainter than VCC941, the former is {\it five} times
larger in effective radius.  Table \ref{data.tab} also lists the total R magnitude,
B--R color, and central B surface brightness derived from the CCD images.

In terms of location within the cluster, the two fields are both close
to the center defined by M87. The ``N'' field is  1.4\arcdeg\ north (or 360 kpc
for a distance of 15 Mpc), and the ``SW2'' field is 1.1\arcdeg\ west (or 290 kpc
distant).  Both are well within the average core radius of the Virgo
galaxy distribution \citep{1987AJ.....94..251B}.

\subsection{ACS Exposures} 
F555W was selected for the V band exposures, primarily because of
our use of that filter in the M81 study, and its small color term
when transforming to standard V magnitude system.  However, it proved to
have substantially lower throughput than the wider and somewhat redder
F606W filter, and the results of the photometry below leave no doubt that the latter
filter would be have been a better choice for this study. However, the
V-I colors provided by the F555W images, along with those in F814W do
allow significant conclusions to be drawn for the brighter member stars.

Each target was exposed on the WFC for a total of 33,280s in F555W, 
and 14,400s in F814W, using 19 orbits per target. The exposure 
details are in Table \ref{data.tab}.  A small dither consisting of
two pointings was used between successive exposures. This allowed
bad pixels to be eliminated in the image combination, but did 
not fill in the gap between the two ACS CCDs, nor did it allow
subpixel sampling.  The galaxies are large enough so that the
loss of sky coverage in the gap is not important.

\subsection{Image Combination} 
The images were taken in the first few months of operation for the
ACS camera, but image combination software was not optimum for some
time, and various unsuccessful reduction
attempts were made following the observations.  
Eventually, the multidrizzle package \citep{2002hstc.conf..337K} which is now a standard
part of the ACS data reduction became available, and was used
to combine the individual 
exposures.  The package rebins the data from the native trapezoidal, 
and spatially variable pixels to square, equal-sized pixels, and as 
part of the combination process can detect and eliminate cosmic rays.
A spatially variable PSF remains, which the model psf used in the 
photometry below takes into account.
The long exposures involved here, particularly for the
F555W images, meant that large numbers of cosmic rays hits were recorded
on the images. These were effectively removed in the combination process.

Possibly because there were only two pointings, there was a fair amount
of pattern noise left in the F555W images. This caused some problems with
the automatic star detection algorithms, which would find false-positives
at higher levels than expected. As well, in one combined F814W image, there was
a small gradient in the background noise level, which similarly caused 
more false-positive stars to be found in the higher noise areas.  These
problems were dealt with by increasing the cutoff signal-to-noise 
slightly, and are of course also taken into account by the artificial
star tests discussed below.

The charge-transfer efficiency of the ACS CCDs is reported to decline with time \citep{Re04}.
The exposures reported here were taken early in the life of the camera, and thus
there is no correction to be applied for that problem.

\subsection{General Characteristics of the Images: Red Giant Stars Resolved in Virgo}
\begin{figure}
\includegraphics[angle=0,scale=0.9]{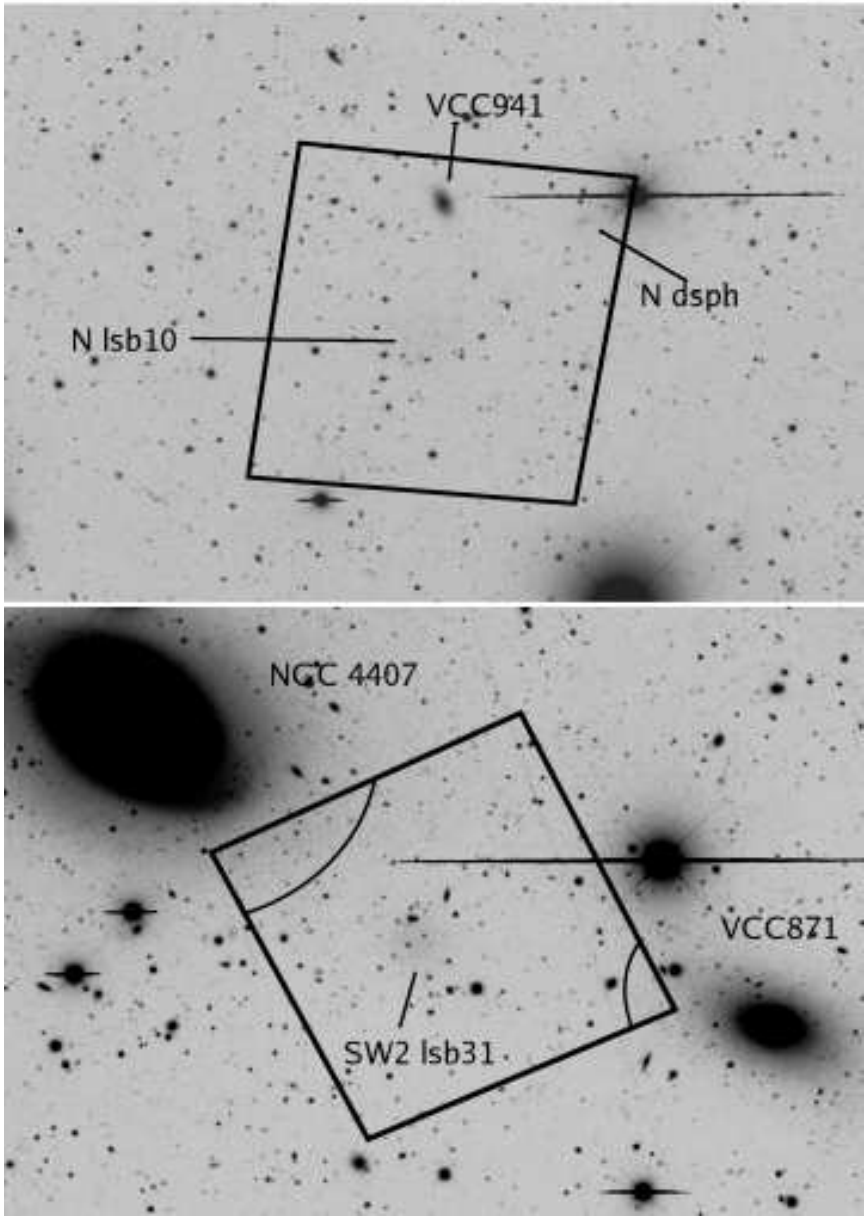}
%\plotone{f1.eps}
%\epsscale{0.6}
\caption{Top: Part of a Mayall 4m telescope plus KPNO Mosaic camera R band image of
the N field. North is at the top; East to the left. The ACS footprint 
(which is about 200\arcs on a side) is outlined, and the three dwarf galaxies
within the ACS field are identified. Bottom: 
Same for the SW2 field. NGC 4407 is a spiral galaxy, while 
VCC871 is a nucleated dE. Halo stars from both galaxies are found on 
the ACS images, whose selected areas are identified by the short, 
elliptical arcs. }

\label{fig:n_field}
\end{figure}

\begin{figure}
\plotone{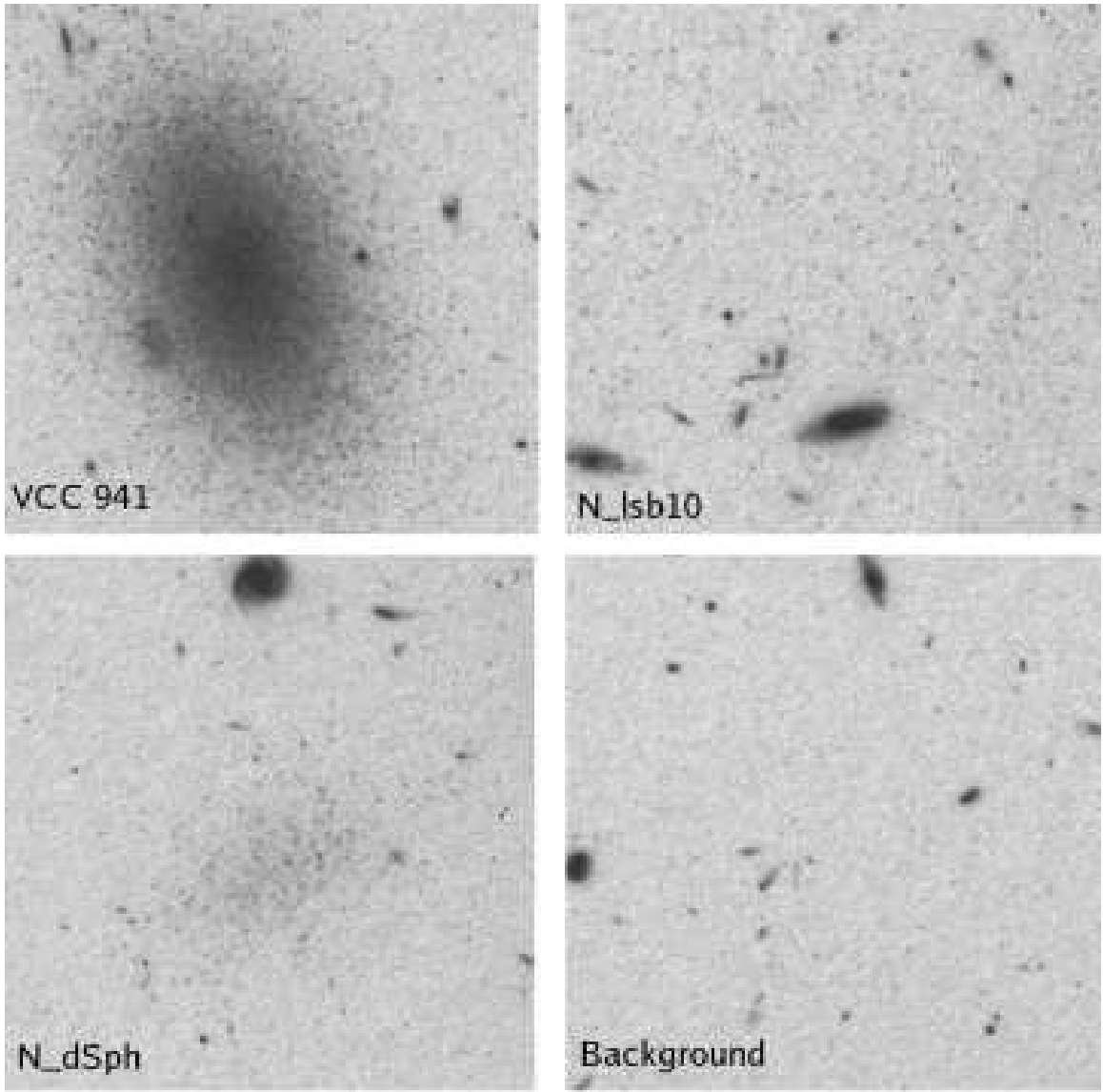}
\caption{The 3 galaxies in the N field imaged by HST/ACS, as well
as part of the field used as background (which also contains resolved
Virgo cluster stars).
North is up; east to the left. Each section is 22.5\arcsec on a side.}
\label{fig:n_close}
\end{figure}

\begin{figure}
\plotone{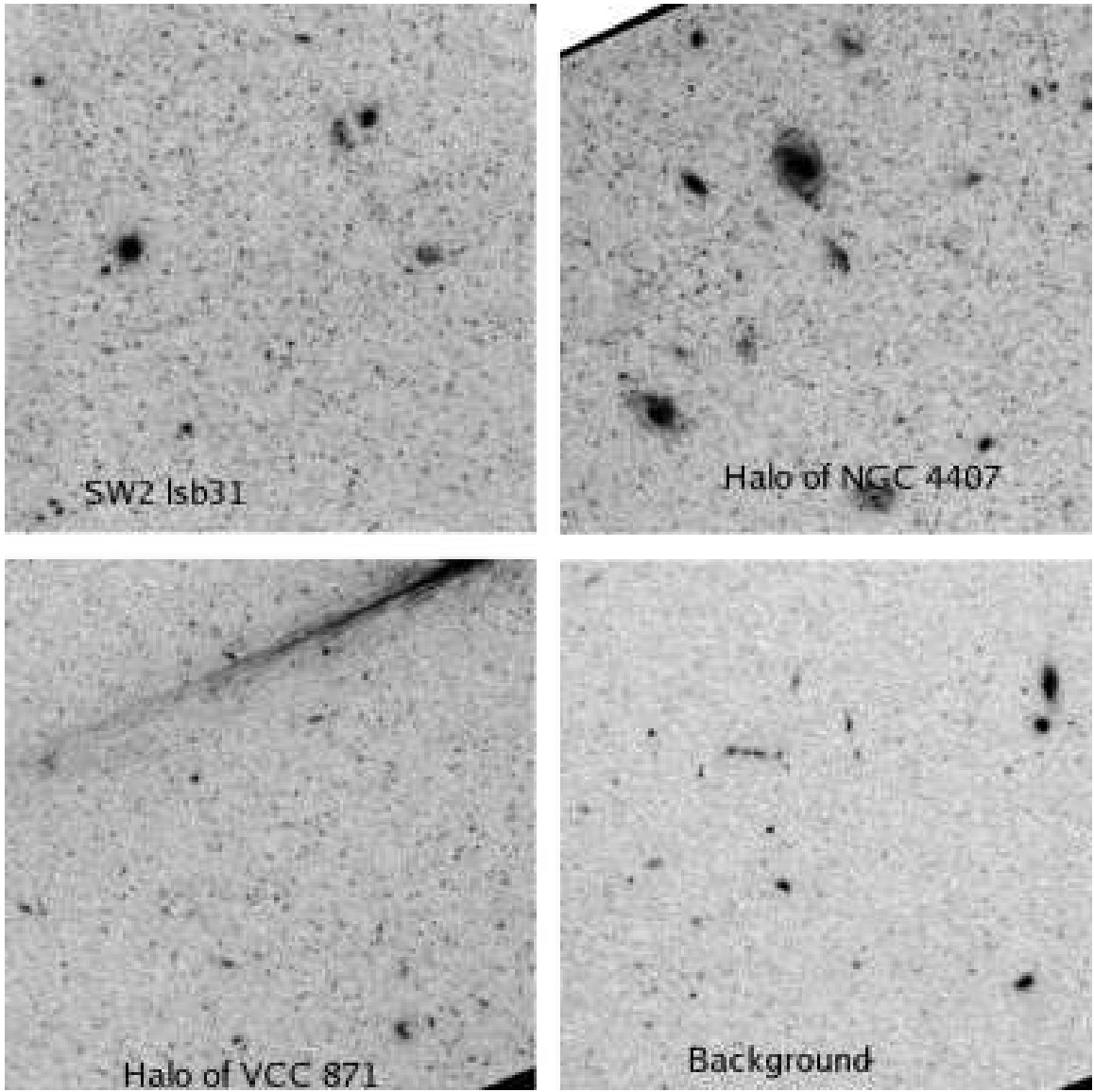}
\caption{The 3 galaxies in the SW2 field imaged by HST/ACS, and 
background field. The flaring in the image of VCC871 is due to a 
nearby, overexposed star.}
\label{fig:sw2_close}
\end{figure}

Figure \ref{fig:n_field}  shows
the ACS footprint on KPNO 4m R band images. As mentioned above,
the outer halos of two galaxies, NGC4407 and VCC871 are within
the ACS footprint for the SW2 field.
Figures \ref{fig:n_close} and \ref{fig:sw2_close} show 
expanded portions of the F814W ACS images of each
of the 6 individual galaxies in the two fields. We now discuss these
in general terms.  The image of VCC941 will certainly come across
best in the reproduction here, and demonstrates dramatically 
that the images do indeed resolve the red giant stars in this
Virgo cluster galaxy. Showing no strong nucleation, VCC941 appears 
in this image
as well-resolved as is the Local Group galaxy Leo I, seen from the 
ground.  A possible globular cluster is seen directly west of the
center. Anticipating the photometric calibrations presented below,
this object has M$_{\rm V}=-6$, and (V--I)$_{\rm 0}$= 1.0, values
consistent with a globular cluster. The diffuse component to the SE 
of the center of VCC941 is probably a background object.

One of the principle targets, N lsb10, has a central intensity 4.5 magnitudes
fainter than VCC941, so it is not surprising that the figure requires
more scrutiny to find its resolved stars. 
As a visual confirmation of the increased star density here, an area near the edge of
the ACS frame is also shown.
Stars in this galaxy in fact extend well beyond the 22.5\arcs diameter of
the figure, going out to a radius of 40\arcsec. The automatic star finding work discussed 
below resulted 
in the detection
of nearly as many stars in N lsb10 as in VCC941.

The third galaxy in the N field was fortuitous, as mentioned already.
It does appear on the 4m CCD images, and was cataloged automatically,
from which we derive its magnitude of R=21.3.  It also clearly
resolves into stars, and combined with N lsb10 reveals the large
range of structures and densities apparent in dwarf ellipticals.

Moving to the SW2 field, the target galaxy SW2 lsb31 also has 
an extremely low central surface brightness, yet stars are clearly resolved
here.  Stars in the halo of the spiral galaxy NGC 4407 are quite dense
in one corner of the frame, where the distance to the center
of the galaxy is 120\arcsec, or 8.7 kpc in projection.  The stars extend 
well towards the center. There is some
overlap with SW2 lsb31, thus a local background subtraction
will be necessary.  In the other corner are found
stars from the dE VCC871, 70\arcs from its center, or 5 kpc in
projection.  It's interesting to note that the velocity of NGC4407 is
102 \kms, while that of VCC871 is 1427. Clearly these two are not
associated, so a determination of their distances would be valuable
in understanding a kinematic detail of the Virgo cluster. Finally, the
density of stars throughout this image which are not apparently 
associated with any of those three galaxies is three times 
higher than is found
in the N field.  Whether these are intracluster stars or 
rather stars tenuously affiliated with the three galaxies may be
addressed by comparing all of the color-magnitude diagrams, as done below.

While the F555W images are not nearly as abundant in resolved stars, 
due to the colors of the red giant stars, there are 
sufficient numbers present to allow colors to be measured for the
brighter giants.

In summary, the objective to resolve red giant stars in Virgo
cluster dwarf galaxies was achieved, and we now describe
the quantitative measurement of the individual stars, and the
cmds which were formed that allowed distances,
and stellar population characteristics to be estimated.

\subsection{PSF Fitting Photometry} 
Stellar photometry was obtained using the DAOPHOT package of
\citet{1987PASP...99..191S}, modeling the spatially variable PSFs 
for each of the 4 combined images separately, using only 
stars on those images.  Multidrizzle supplies an image that
has been background subtracted and scaled to a 1s exposure time,
so the first step was to rescale by the exposure time of 
a single exposure, and add back in the average background level
for each set of co-averaged exposures.  

PSFs were constructed using 15-20 bright stars which had no pixels
above a level of 20,000 counts, the point
at which an ostensible non-linearity set in and the PSF no longer
matched those of fainter stars.  The PSF model had a
quadratic dependence on location. Given the 0.05\arcs pixels
and the resultant under-resolution of point sources, the PSFs
created do not perform as well for the ACS data
as do PSFs for better-resolved data,
but such errors should
be taken care of in the aperture correction step.
The PSF was defined over 25 pixels
(there is a large halo in the F814W PSF, due 
to scattering, see \citet{2005PASP..117.1049S}), but was fit within a radius of
2 pixels.  It was normalized to aperture photometry within 10 pixels
(0.5\arcsec).  Aperture corrections were also measured using these stars,
to determine any photometric offset between the psf photometry and
the aperture magnitude within 0.5 \arcsec.   \citet{2005PASP..117.1049S} have
provided aperture corrections from that aperture size to infinity, 
in all ACS filters.

The DAOPHOT automatic star detection algorithm was used to collect
star centers for the PSF photometry.  Since we attempted to 
measure stars at the detection limit, any such method will be 
sensitive to the noise characteristics.  After some testing,
a threshold of 3.3 sigma was used for detection of stars in the
F814W images. This avoided
creating too many false-positives over areas of the detector
which had higher noise than average, but still cataloged nearly all
the stars in the target galaxies, as determined visually.  Since
the F555W images are not as deep, and have a remnant pattern noise,
the star catalogs from the F814W
frames were used as input for the F555W work, rather than creating
a separate F555W catalog and matching 
stars at a later step.  This technique meant
that there are stars with valid photometry that are significantly 
detected in the F814W data, but which are well below the F555W limit.
For those stars,  we have in effect measured the I band magnitude, but only a blue
limit on its V--I color.  The uncertainties for the photometry as well as
the artificial star tests were then used to guide the analysis so that
such stars are treated properly in the color-magnitude diagrams.

Generally, two passes of photometry were run. First, a star list was made and
entered into ALLSTAR, which aside of the photometry, produces a star-subtracted
image. Stars missed in the first round were located in the subtracted image 
and added to the original list.  The original frame was then measured again
by ALLSTAR.

The star catalogs with the F814W and F555W photometry were then
cleaned of several kinds of bad measurements due to
misidentifications: bright star
halos, bleed columns, diffraction spikes and parts of bright background galaxies.
SExtractor \citep{1996A&AS..117..393B} was run on the F814W images to provide classifications of objects
down to about 1 mag above the limit. Those classes 
were then used to clean fainter galaxies
out of the catalogs, as well as close stellar pairs.  
Remaining stars with a chi squared fit to the PSF function greater 
than 1.0 were also deleted.
The photometry was then placed on the standard Johnson/Kron-Cousins
VI system using the aperture corrections and synthetic transformations 
provided in \citet{2005PASP..117.1049S}, Appendix D, tables 5 and 18.
Some care has to be taken as the transformation equations used depend on 
instrumental color. Coefficients and aperture corrections were taken from Tables 22 and 5 
of \citet{2005PASP..117.1049S}. The transformation uncertainties comprise those from:
the aperture correction from the DAOPHOT aperture to 0.5\arcs,  the 
correction from 0.5\arcs to infinity, and the the transformation
to the standard system.  In the I system, these uncertainties are 0.007, 0.002 and
0.024 mag, respectively for a total of 0.025; for V, they are 0.014, 0.001, 
and 0.01 for a total of 0.017.  These are small compared to the uncertainties
in the average tip magnitudes and colors used below.

Radial limits were defined for each of the 6 galaxies, following their average
elliptical isophotes, thereby allowing stars from the full image catalogs to be 
associated with particular galaxies, or a local background.
The limits were chosen to be approximately at the radius 
where the star density was about twice that of the ``background,'' where the latter
is estimated from areas on the frames most remote from the 6 galaxies. The center
of VCC941 was sufficiently crowded to affect the photometry, so stars
within the elliptical area with geometric mean radius of  3.5\arcs were also excluded.
For
the two large galaxy halos that appear in the corners of SW2 field, the areas
are of course wedges, since the galaxy centers are far off of the CCD images
(outlined in figure  \ref{fig:n_field}).

The background area on the N field is the southwest corner of the image.
For the SW2 field, several background areas were chosen - one surrounding 
lsb31 itself, used in correcting the luminosity function and cmd for 
contamination from NGC4407, another in the southeast corner, and a third north of 
VCC871. The latter two were used to investigate the question of the extent
of NGC 4407.

\subsection{Artificial Star Tests} 
DAOPHOT calculated and listed photometric uncertainties for each star,
using photon statistics in the apertures, the uncertainties
in sky determination, and the readout noise of the CCD detectors.
Mean values of the uncertainties in I and V--I were then computed for 0.5 mag bins
in I magnitude along the apparent giant branches.  At I=27.0, the apparent location
of the giant branches, the mean uncertainty in color is about 0.19 mag. 
At I=28.25 mag, the uncertainty has increased to 0.3 magnitude.   The cmds
below show all the calculated uncertainties in graphical form.

It is typical in color-magnitude diagram work to compare such internal estimates of
the uncertainties with those derived from separate observations of the same field.
The present data set doesn't lend itself to such analysis because most of the
stars are near the detection limit, and producing two separate images instead of
one from the total frames available would render many stars difficult to measure.
However, artificial star tests may be performed, which give an additional
estimate of the uncertainties.  An artificial giant branch was created, using
the mean colors and magnitudes estimated from the SW2 lsb31 cmd.  Stars with those
I and V magnitudes were randomly placed in the SW2 F814W and F555W images (after
converting the photometry back to instrumental values), and the images were processed
as before.  Each pass in the process consisted of adding 700 stars to the images; 
passes were run until the statistics no longer improved. Differences in input and derived magnitudes and colors were collected,
and means formed.  These mean uncertainties are also shown in figure  \ref{fig:sw2_cmd} and
are essentially the same at each magnitude bin
as those found from the method using the calculated photometric uncertainties on the
real stars.

Artificial star tests were also performed to assess the limiting
I magnitude,  and limiting V--I color as a function of I 
magnitude, important quantities in the analysis of the color
magnitude diagrams in that the reddest giant stars will
likely be fainter than the detection limit in the F555W images.
At each 0.25 magnitude bin between magnitudes 27 and 29.5
for both filters, 100 stars were added at random to the images.
The photometric reductions were run again as above, to 
determine what fraction of the added stars were recovered. This
was done 5 times at each magnitude bin, to average out the 
results. The result was that 50\% of the stars are lost at
I=28.2, and at V=28.95.

\section{Color-Magnitude Diagrams}
\begin{figure}
%\plotone{f4.eps}
\includegraphics*[angle=0,scale=0.8]{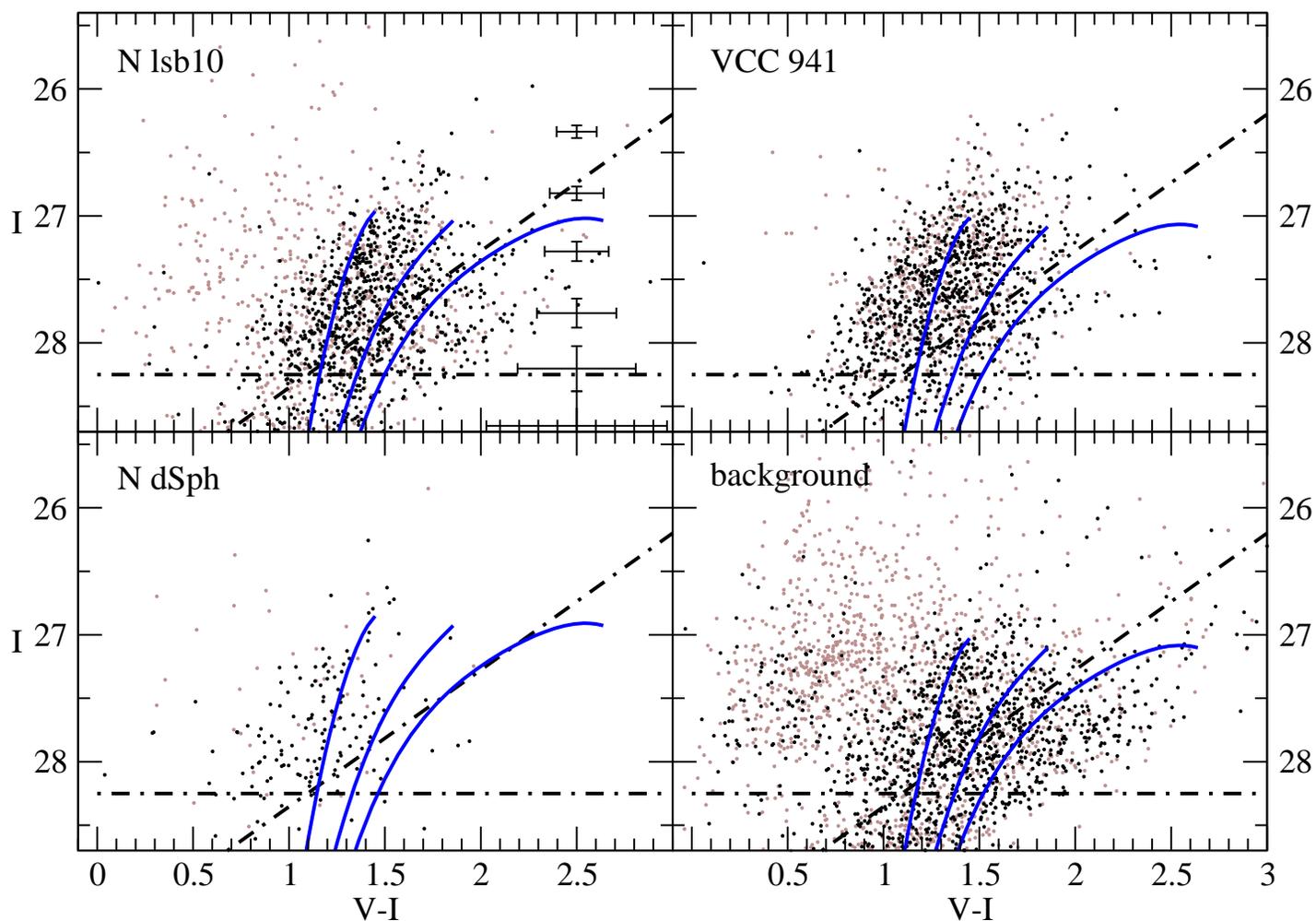}
\caption{VI cmds of the N field. MW globular cluster giant
branches are shown for comparison: M15 ([Fe/H]=-2.17), NGC1851 (-1.16), 
and 47Tuc (-0.71).  These have been corrected to the derived distance modulus
for each object, and for the assumed extinction.  Faint dots represent 
objects classified as non-stellar, which includes some close stellar pairs
near the centers of the target galaxies,
but mostly background galaxies.
Average uncertainties in I and V-I, derived from DAOPHOT, are shown for the lsb10 field, at specific magnitudes.  
Dot-dashed lines represent the limit where 50\% of the stars in the 
artificial star tests are not recovered. The area used for the cmd
of lsb10 is 28\% of that of the background field, while VCC941 and N dSph are
both 2\% of the background area. }
\label{fig:n_cmd}
\end{figure}

\begin{figure}
\includegraphics*[angle=0,scale=0.8]{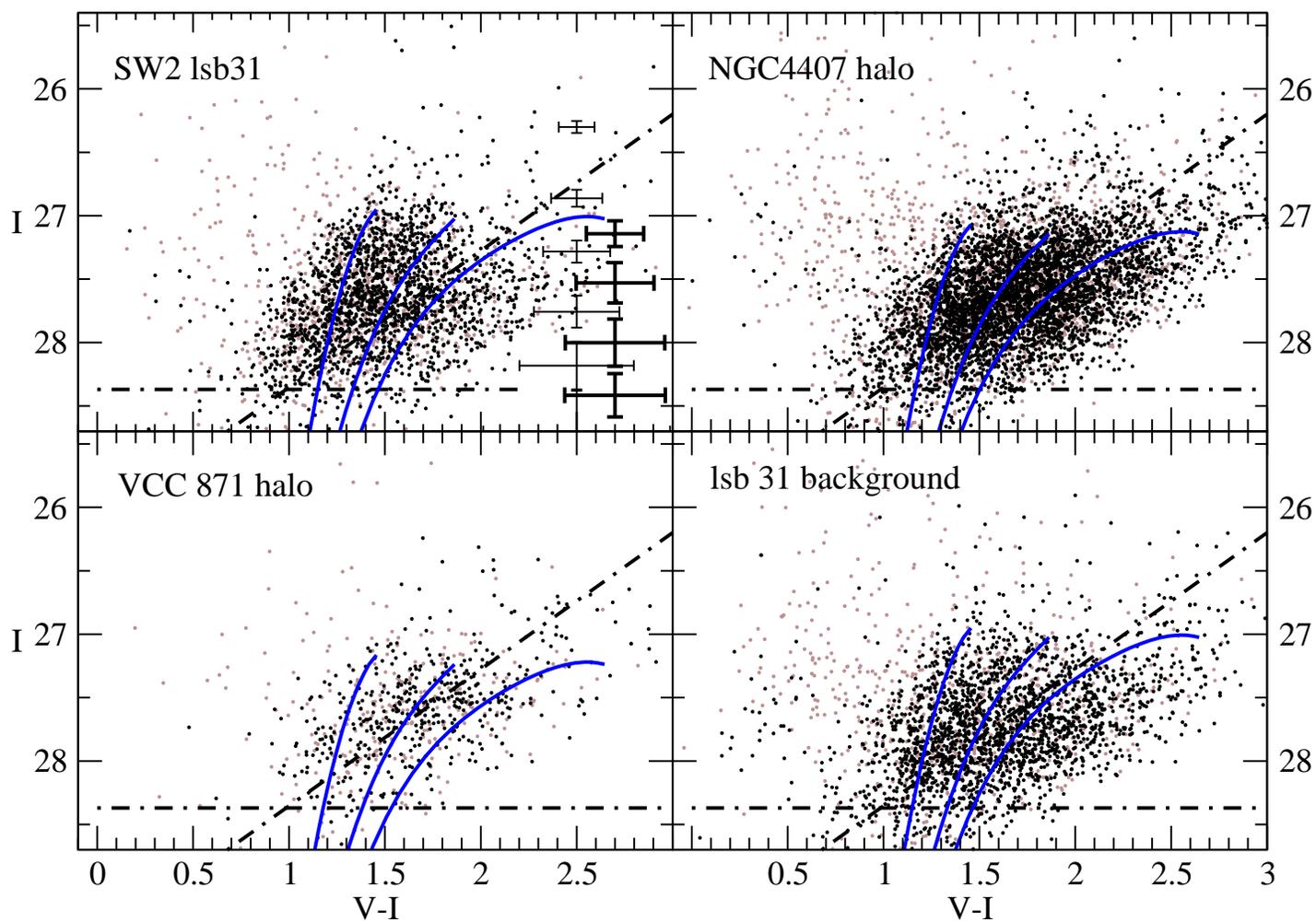}
%\plotone{f5.eps}
\caption{VI cmds of the SW2 field. Average uncertainties in I and V-I are shown for the lsb31 field.  The heavy lines show uncertainties determined from the artificial star
tests, which are largely in accord with the uncertainties provided directly by
DAOPHOT.  The lsb31 background is the area 
on the sky surrounding lsb31 that was used for the luminosity function 
subtraction.  The giant branches are as in figure \ref{fig:n_cmd}, 
except for the
background field where they have been
copied from those shown in the lsb31 cmd, for comparison purposes.}
\label{fig:sw2_cmd}
\end{figure}

\begin{figure}
\includegraphics*[angle=0,scale=0.8]{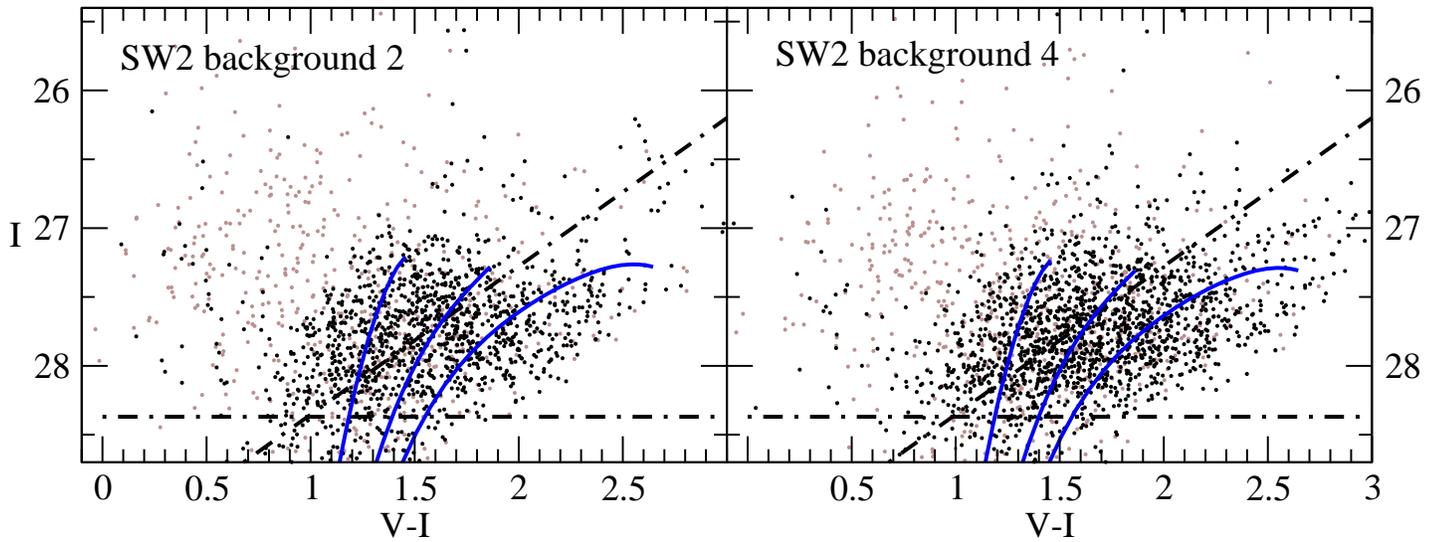}
%\plotone{f6.eps}
\caption{VI cmds of two different background areas in the SW2 field. The area
incorporated in the background 4 field is 1/3 that of the N background field.}
\label{fig:sw2_cmd2}
\end{figure}

Figure \ref{fig:n_cmd} shows
the color-magnitude diagrams for the N field, 
while figures \ref{fig:sw2_cmd} and \ref{fig:sw2_cmd2} show
the cmds for the SW2 field
The average uncertainties 
in color and magnitude described above 
are shown only in the N lsb31 and SW2 lsb31 cmds, though
they apply to all the graphs in each figure.  The error bars with 
heavy lines represent the uncertainties determined by artificial star
tests.
The limit for 50\% recovery in I is
shown as a dot-dashed line at I=28.2, and the diagonal line
incorporates that limit with the one in V to arrive at 
the color limit.  Faint dots are objects classified as
non-stellar by SExtractor, though many of these are in fact
crowded stars which DAOPHOT was able to obtain satisfactory
photometry, but which SExtractor did not separate into
individual stars.  Nonetheless, these are excluded from the
derivation of metallicity and distance.

There are two obvious results from these diagrams. First,
giant branch stars have been successfully detected, down to
perhaps 1.2 magnitudes below the tip. Second, there are
strong incompleteness factors for stars on the red side of 
the giant branches, stemming from the insufficient depth
of the F555W exposures.     The fiducial giant branches of 
galactic globular clusters are shown in this figure (see below),
which indicate that at least $\sim 50$\% of stars in  
a 47 Tuc population would be missed in 
the F555W filter (though such stars would show up in the F814W 
images).  An even larger fraction of stars in a more metal 
rich population would be lost. Such an incompleteness would
thus artificially lower estimates of the metal abundance
in metal rich populations.   For the four faint galaxies though,
N lsb10, VCC941, N dSph, and SW2 lsb31, the cmds show a distinct
gap between the distribution of detected stars and the 50\%
loss line, and so we expect the systematic effects in the
metal abundance determinations to be small.  To verify this
assertion, we studied the stars found on the F814W frames but not
on the F555W frames.  If a large number of red stars
had been missed in the cmds, the brighter ones should still have
appeared in the F814W frames and been cataloged.  For the
apparently reddest population, that of NGC4407, indeed about
6\% of the stars detected in the F814W frame brighter than I=27.5 
had no F555W measurement, and are missing from the cmd. However,
for SW2 lsb31 for instance, only 2\% of the brighter stars are
missing from the cmd.

In comparing the cmds for the different objects and
backgrounds, it seems
clear that all these have tip magnitudes close to I=27, 
meaning they are all at about the same distance.
N dSph appears to be more distant by about 0.2 mag, though
the paucity of stars (not surprising given its
luminosity of V=--9.6) makes the tip determination especially
uncertain.  Its giant branch
is distinctly 
the bluest one of those presented here.  A test confirming
that was performed by selecting stars randomly from the cmd
of VCC941 in numbers equal to the number of stars
in the N dSph cmd. Out of 100 tests, no sample from VCC941
had the same color distribution as the stars in N dSph.

While the mean uncertainty in color at the TRGB is around 0.19 mag, the rms scatter
in observed V--I colors is around 0.22 for N lsb10 and VCC941, and 0.24 for
SW2 lsb31.  This may mean that a spread in metal abundance has been detected,
though for SW2 lsb31, the spread is at least in part due to the presence of
RGB stars from NGC4407.  The 0.3 mag uncertainty in V--I at I=28.25 explains the large
spread in color at the fainter magnitudes.
The redder giant branches of VCC871, NGC4407 and the two background 
fields
will naturally mean that only lower limits to their 
metallicities could be estimated
with the current data.

Upper AGB stars indicative of an intermediate age population
are occasionally present in metal-poor populations,
such as the dwarf elliptical 
F8D1 in the M81 group \citep{1998AJ....115..535C}.
If present in these Virgo galaxies, they would extend
over a magnitude brighter in I, and range in color from that
of the blue edge of the giant branch tip to several magnitudes
redder (e.g., $ 27>{\rm I}>26; 1.5<{\rm V-I}< 4$). 
There is some evidence for these in the cmds of each of the four 
low luminosity galaxies, though
the numbers are small,  and a quantitative analysis of the 
fraction of light contributed by an $\sim 8 $ Gyr population
isn't warranted by the data. A qualitative comparison with the 
two M81 dwarf galaxies studied in \citet{1998AJ....115..535C}, concludes
that no more than 30\% of the light in the Virgo dwarfs could
be from an intermediate age population.  The RGB tip of NGC 4407
may be confused by the presence of stars more luminous
than M$_{\rm I}=-4$, such as are found in populations as metal
rich as 47Tuc. The halo cmds of the  nearby spirals observed with
WFPC2 by \citet{2005ApJ...633..810M} also show a number of stars above the
RGB tip.  

A small number of  blue stars appear in these cmds, with
V--I colors bluer than 0.7.   About half of these are clearly non-stellar
on inspection, objects which the SExtractor classifications indicated 
an uncertainty.  A few are clearly stellar though, and 
we point out one star in particular in VCC941, 
seen at I=27.3 and V--I=0.06 in 
figure \ref{fig:n_cmd}.  The numbers of the apparently real blue stars
are too low to indicate
recent star formation; rather they may be examples of
post-horizontal branch stars (``UV-bright stars'' \citet{1974ApJ...193..593Z} ),
though the M$_{\rm V}=-3.5$ for the VCC941 stars is brighter than 
any known UV bright star in galactic globulars. Other
photometric colors would be helpful in sorting these stars out,
though the  V--I colors already  rule out planetary nebulae.

\section{Distances and Metal Abundances}
The tip of the red giant branch represents the luminosity at
which the helium core flash begins, a value that varies slowly with
age and metal abundance for old, metal-poor populations 
(Da Costa and Armandroff 1990).  The technique of deriving 
distance modulii by measuring the
I band magnitude of the tip, converting to a bolometric luminosity,
and then using a calibration of tip luminosity with abundance has
become a common tool \citep{1993ApJ...417..553L,1998AJ....115..535C,
2003A&A...404...93K}.  We also estimated the metal abundances using
the color of the giant branch, using the quadratic calibration provided 
in \citet{1993AJ....106..986A} and \citet{1998AJ....115..535C} . 
The process of deriving distances and metal abundances 
was thus iterative in nature.
Reddenings were taken from the catalog of \citet{1998ApJ...500..525S}, and 
are listed in Table \ref{res.tab}.

\begin{figure}
\vskip 1.0truein
\plotone{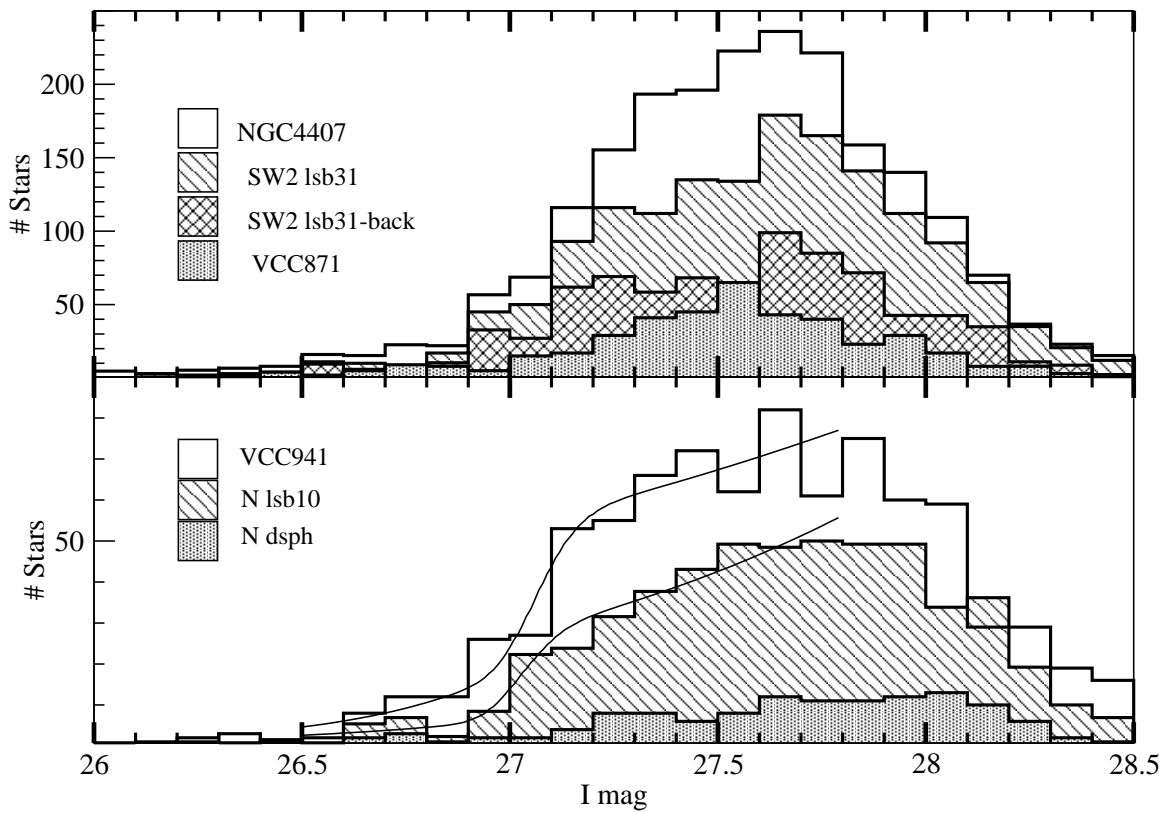}
\caption{I band luminosity functions for the three galaxies in 
each of the N and SW2 fields. Magnitudes were binned into 0.1mag bins, excluding
extremely red and blue stars. The areas used were not equal for
the galaxies, see the text. In the top panel, the galaxy SW2 lsb31
is shown twice, both with and without a local background subtracted.
For display purposes, the histograms of N lsb10 and NGC4407 have been divided by
1.3 and 1.5 respectively.  The best fit analytic luminosity functions are shown
as solid lines in the lower panel for N lsb10 and VCC941.}

\label{fig:both_lf}
\end{figure}

We began by forming I band luminosity functions from the stellar
photometry catalogs.  The background star density in the N field
is low enough to ignore for the three galaxies in that field, 
N lsb10, VCC941, and N dSph, but such is not true for the SW2 field.
An annular region around SW2 lsb31 was chosen to correct the function
for that galaxy, though in the end, the correction did not affect
either the tip magnitude or the derived distance and metallicity.
Stars with colors 0.3 mag redder than the color cutoff line
were excluded, as were very blue stars. Magnitudes were put into 0.1 mag
bins, and those are shown in Figure \ref{fig:both_lf}.  Only the histogram
for SW2 lsb31 has been background subtracted. The histograms
do not represent the same area on the sky, so the relative scales
among galaxies in the figure  are unimportant.

\begin{figure}
\includegraphics*[scale=0.7]{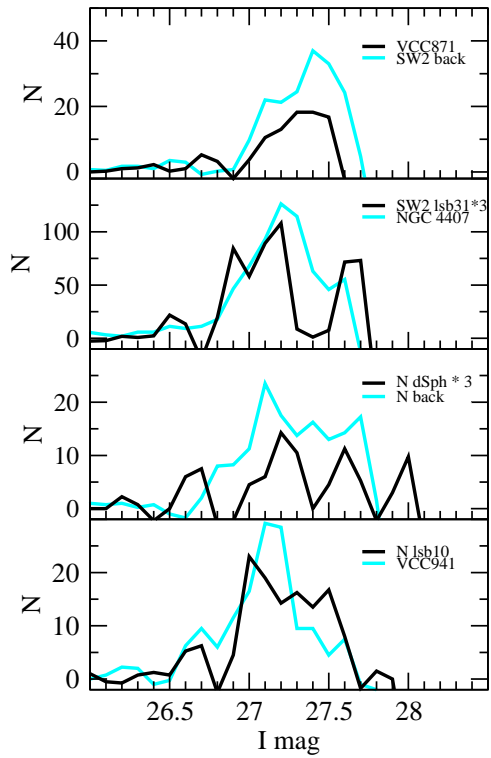}
%\epsscale{0.1}
%\plotone{f8_color.eps}
\caption{Sobel-filtered I band luminosity functions for the six galaxies and
two background populations.}
\label{fig:sobel}
\end{figure}

All of these histograms show a clear slope change near I=27, with
varying degrees of significance. N dSph is probably the weakest
case, but the cmd shows pretty convincingly that the tip is near
I=27 as well.  
The apparent tip magnitudes were
estimated using two techniques. First the binned luminosity functions were
passed through a Sobel filter (-1,0,+1) \citep{1993ApJ...417..553L} , 
with uncertainties set by choosing
different binning parameters, which are found to be 0.1 mag 
for all but N dSph and VCC871, which are 0.15 mag.  The filtered 
luminosity functions are shown in figure  \ref{fig:sobel}.  While these
do indicate that the gb tips are clearly within 0.2 mag of I=27,
the functions for several objects are noisy, and the slopes are not
very steep, due to the small numbers of stars. 

A maximum likelihood method was used as a second estimator of the 
apparent tip magnitudes, and was patterned after that described in
 \citet{2002AJ....124..213M}. In brief, the luminosity distributions were
represented 
with a broken-power-law function, with the break between the powers occurring at the RGB tip.  
The observed errors at each magnitude were used as a smoothing function
for the power-law.
A grid of likelihoods was formed by varying the 4 parameters of the function:
the two powers, the 
magnitude difference between the powers, and the tip magnitude.  The parameter
set giving
the maximum likelihood was identified, which gave the most probable
value for the tip magnitude.  Uncertainties in the tip were calculated by a bootstrap
monte carlo method, whereby the star sample for each galaxy is randomly
sampled, with replacement, many times.  The standard deviation of the tip magnitude
values from the all the runs is then taken as the uncertainty in the measurement of
the tip. Several of the derived analytic luminosity functions
(including the error smoothing) are shown along with the binned luminosity 
functions in figure \ref{fig:both_lf}. 

Both of these estimators have drawbacks - the edge detection method is
sensitive to the binning, while the results from the maximum likelihood method 
depend on the parent analytic function being correct. However, the two estimators 
generally gave tip magnitudes within the associated errors, with 
the maximum likelihood method tending to average the noisy peaks seen
in \ref{fig:sobel}.  The maximum likelihood values are listed in 
table \ref{res.tab}.

To
derive a distance then, we used an iterative process that incorporates the
dependence of bolometric correction on the color of the giant
branch tip (BC$=0.881 - 0.243{\rm (V-I)_{tip}}$), 
the dependence of bolometric luminosity of such stars on [Fe/H] 
( $\rm{M}_{\rm bol}=-0.19[Fe/H]-3.82$, both from  \citet{1990AJ....100..162D}), and the observed relation between [Fe/H] and mean giant branch color
at M$_{\rm I}=-3.5$ (about I=27.5 here, [Fe/H]=$-1.00+1.97\rm {(V-I)_{-3.5}-3.2(V-I)}_{-3.5}^{2}$   from \citet{1993AJ....106..986A} ).
With this method, we derived the distances and [Fe/H] estimates for the six galaxies as
well as three different background areas in the images, which are listed in 
Table \ref{res.tab}.  One can also avoid the dependence of  $\rm{M}_{\rm bol}$ on
metallicity, and simply use  $\rm{M}_{\rm I}$, which is nearly constant at -4.05 over
the metallicity range likely for the dwarfs. The resultant
modulii are not significantly different from the values in the table.

The uncertainties in distance modulii were derived
from the photometric calibration uncertainties ($\sim 0.025$mag), the uncertainty in extinction
(0.02 in E(B-V)),  the  uncertainty in determining the 
RGB tip (0.10-0.15mag), and the uncertainty in  $\rm{M}_{\rm bol}$ induced by the
uncertainty in [Fe/H] ($\sim 0.015$mag).
Likewise, the uncertainties in [Fe/H] were derived by propagating the uncertainties in
tip magnitude determination, uncertainty in the measured V--I color of the stars, 
and the photometry uncertainties, and then
all combined with the calibration uncertainty of 0.08dex. 

These distance modulii along with the reddenings
were used to place fiducial galactic globular cluster giant branches in 
figures \ref{fig:n_cmd},  \ref{fig:sw2_cmd} and \ref{fig:sw2_cmd2}.
The location of observed stars with respect to the globular sequences
supports the tabulated distance modulii and abundances (and demonstrates
the limitations on the [Fe/H] results for the more metal rich halo fields and
the background fields).

\section {Distance to Virgo Using the RGB Tip Stars}
While there are strong biases in the  [Fe/H] determinations
for the two halo fields and the backgrounds, the distance modulii
values should be relatively unaffected by the redder giant
branches.  The color term in the conversion  of the ACS F814W
magnitude to standard I for the red giant stars is $-0.014(V--I)+0.015(V--I)^2$.
and thus systematic errors of only 0.03 mag would exist for stars 
with estimated  (V--I) $\approx 2.0$, but which have
true (V--I) $\approx 2.5$, in the sense that the
estimated magnitudes are too bright. This potential error appears
small enough to ignore.
Thus we calculated a distance for the Virgo cluster
using intracluster stars as well as distinct galaxies.  The mean of
the distance moduli for the 6 galaxies  observed here is  $31.0\pm0.05$, for 
a distance of $16.1\pm0.4$ Mpc.
The distance modulii for the intracluster stars in both fields
is $31.2\pm0.09$  ($17.4\pm0.7$ Mpc).

These distances compare favorably with the value of  $15.7\pm1.5$ Mpc of
\citet{1998Natur.395...45H}, and also
with the distance $16.1\pm1.0$ Mpc  found by the
HST key project group \citep{2000ApJ...529..768K}, using Cepheids found in 5 Virgo spirals, and calibrated
with LMC P-L relation, where the LMC distance modulus was taken to be 18.50.
The distance scale used here (from \citet{1990AJ....100..162D}, but originating 
in \citet{1984ApJS...55...45Z} ) is tied to
galactic globular cluster distances which use an RR Lyrae 
scale with M$_{\rm V}$=0.82 + 0.17[Fe/H] \citep{1990ApJ...350..155L}.  Given
the long history of discrepant distances to Virgo, it is
gratifying 
that the distance here, taken from a method using old populations, agrees
well with the method that uses young populations.

\citet{2006astro.ph..3647S}  reported a Virgo distance of $19.7\pm1.5$ Mpc 
(m--M=$31.47\pm.16$), from the
combination of Cepheid distances to 4 galaxies, SN Ia luminosity 
distances to three galaxies, and the Tully-Fisher relation for 49 spirals.  Such a large
distance estimate is clearly incompatible with the magnitudes of the red giant 
tip stars found in any of the galaxies or for the intracluster stars presented here. 
However, five of the seven individual galaxies in that study have distances in 
accord with the new RGB tip distance to the cluster.

\subsection {Virgo Cluster Depth}
The exceedingly small areas imaged here with ACS preclude any
meaningful measurement of the cluster depth, but a few items
are worth discussing.  There is indeed a significant detection
of differences in distances among the galaxies, at the level of
0.2 mag, or 1.5 Mpc.  This corresponds well with the depth
estimate of 1.2 Mpc found by \citet{1990ApJ...356..332J} from Virgo PN, and within the upper limit
of 3.4 Mpc found by  \citet{2005ApJ...634.1002J} from the globular cluster luminosity functions
of Virgo early type galaxies.

Second, the intracluster stars appear to rule out
a distance (at least for those areas of the cluster)
smaller than 15 Mpc (m--M=30.9).  
A large depth of the cluster would diminish the slope of the
field star luminosity function at the red giant tip.  The general effect was
simulated using an observed I band luminosity function derived from
ACS/HST data of an M81 dE galaxy \citep{M81prep}, shifted to the distance modulus
of Virgo.  The luminosity function was then distributed along the line
of sight using a  gaussian $\sigma=2$Mpc (0.23 mag), and inspected.
Such a depth would have washed out the observed I band rise for the 
N field, and thus the Virgo depth must be smaller than 2Mpc.  
A depth corresponding to  $\sigma=1$Mpc (0.13 mag) is
more in accord with the observations here, though clearly this is a topic
for which more deep ACS fields at different positions in the cluster are
needed to provide a definitive result.

\subsection{Intracluster Stars} 

\begin{figure}
\includegraphics*[scale=0.7]{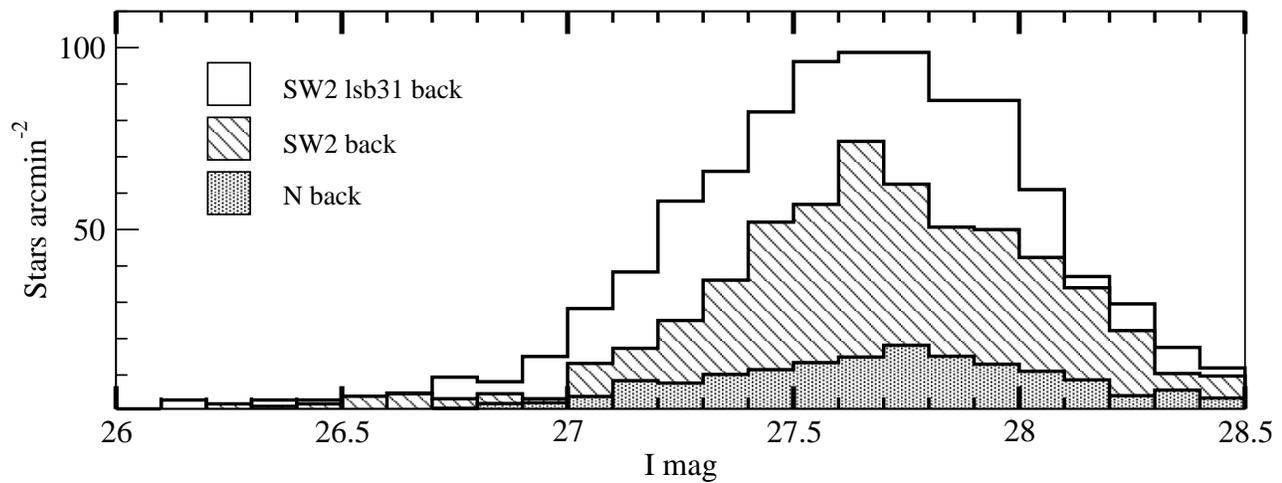}
%\plotone{f9.eps}
\caption{I band luminosity functions for the background fields, 
in stars  arcmin$^{-2}$.
Two are shown for the SW2 field, one surrounding SW2 lsb31, and 
the other in the lowest density region of the image. Note the
much higher density of stars in the SW2 field, as well as the break
in the function at around I=27, meaning that free-floating giant stars have 
been detected in Virgo.}
\label{fig:back_lf}
\end{figure}

The luminosity functions for the background fields 
are shown in Figure \ref{fig:back_lf}. Here the
histograms represent stars arcmin$^{-2}$ , and one can see the large
difference in star densities between the N and SW2 field.  Remarkable
is the fact that even in the N field, where there are no large
galaxies nearby, and the three target dwarfs have well defined limiting
radii,  a clear break in the function for the background
is seen also at I=27, meaning 
that free-floating red giant branch stars have been detected in the
Virgo cluster.

Some
information may be obtained about the metal abundances
of the N field intracluster population by comparison
with the other cmds,  bearing in mind the limitations
of the present data and assuming that the color limits
are the same for all the objects.
For N intracluster stars 0.5 mag fainter than the tip,
the mean V--I color is 1.45, with an rms of 0.26 mag.
That is somewhat bluer than the mean color of 1.67 for
the halo of NGC4407,  meaning the N field stars are more
metal poor.  The rms is larger than that found in the three
well observed dwarfs (N lsb10, SW2 lsb31, and VCC941), 
thus we can also conclude that there is a larger spread in 
abundances than found in the dwarf galaxies.

\begin{figure}
\includegraphics*[angle=0,scale=1.0]{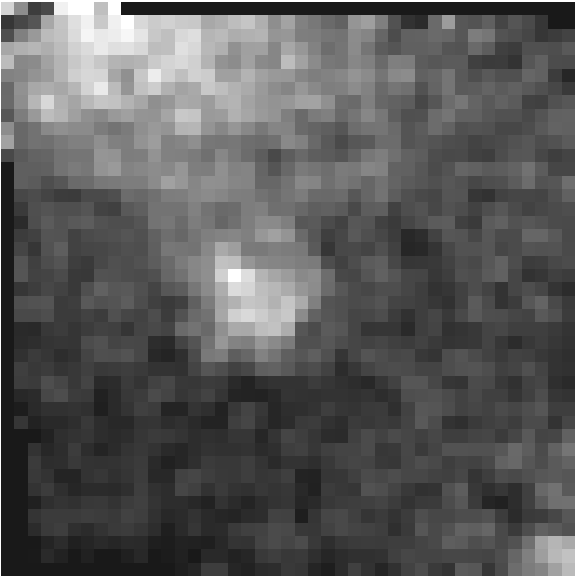}
%\epsscale{0.5}
\caption{Density of Resolved stars in the SW2 field F814W image. North is at about 1 o'clock,
and the field diameter is that of ACS, 200\arcsec.
The central object is SW2 lsb31, with a peak density of about 400 stars arcmin$^{-2}$,
where the limiting magnitude used is I=28.3.
At the upper left is the halo of NGC4407, while that of VCC871 is at the lower
right. The lowest density of Virgo RGB stars in this field is found at the lower left.}
\label{fig:sw2_stars}
\end{figure}

The SW2 field stars are possibly related to the spiral NGC4407,
though the distribution is not continuous.
Figure \ref{fig:sw2_stars} shows the distribution of resolved stars 
in SW2 with I $> 28.3$; the image has been binned substantially to
increase the contrast.  
The SW2 background field was taken from the lower left of this image,
which is around 23 kpc radially distant from the center of NGC4407.
The halo of NGC4407 clearly extends to the
center of the image, where lsb31 lies, but not obviously beyond that.
Yet, the distance moduli of the spiral and that of the background
are similar, and distinct from lsb31.  A case can be made that the
SW back2 stars are in the mean more metal-poor than  those of
the spiral halo, though that may be expected given the galactocentric
distance.

A rough
conversion to surface brightness for the Virgo stellar background
can be made by using I band surface
photometry and the ACS resolved star photometry in the F814W image for
VCC941 collected into the same radial bins. A mean V--I was measured for
the galaxy (0.9), given a conversion from star counts in F814W to V 
surface brightness. Applying this to the background star counts, we derived a
surface brightness for the N background stars of $29.0 \pm 0.5 $ mag arcsec$^{-2}$,
in V, and 27.0 for the SW2 field.  (The conversion can also be used to
calculate central surface brightnesses of the two extremely low surface
brightness galaxies, N lsb10 and SW2 lsb31, which can't be measured with
surface photometry of the ACS images. These are compared with the 
KPNO 4m data in table \ref{res.tab}.)

Both ACS fields are hundreds of kpc from
from M87, which reaches V=29 mag arcsec$^{-2}$ at about 0.4 degree radius, or 130 kpc \citep{1976ApJ...209..693O},
thus the N background population is likely not part of the extended halo of that cD
galaxy. It is comprised of true intracluster stars.  
The background surface brightness for the N field is 
lower than estimates made from PN surface densities
\citep{2004ApJ...615..196F}, which range from V=26.5 to 28.4.  The PN 
estimates require a number of assumptions to be made about the PN
parent populations, which are probably more uncertain than the corrections
made here (a constant V-I was assumed here, as well as a constant detection
bias between the calibrating galaxy VCC941 and the background fields).
In any case, the surface brightness reported here of V=29.0 refers to a
single location in the cluster, that of the N field.

\section{Comparing Extremely Low Surface Brightness Galaxy Populations } 

The mean metal abundance for dwarf elliptical galaxies has been
known to correlate well with the luminosity of the galaxy, a fact
attributed to the onset of mass loss in galaxies where the potential
well is not deep enough to retain the mass lost from evolving
stars \citep{1998AJ....115..535C, 1986ApJ...303...39D}.  The Local Group dwarf spheroidals define a fairly tight relation,
when the abundances are all determined via the same method (RGB tip color),
though there is some scatter at the faint end, and only a few galaxies
define the bright end.  Dwarf ellipticals in the M81 group also exhibit
this relation, and furthermore show that surface brightness is not as
well-correlated with abundance as is luminosity.  A luminous (for a dwarf)
galaxy that has very low surface brightness in the group (M81-F8D1) has
an [Fe/H] value that is in agreement with Local Group dwarfs of similar
luminosity, but in disagreement with those of similar surface brightness.
The two extremely low surface brightness galaxies observed provide
a further test of the relative importance of total luminosity over surface
brightness in determining mean abundance, and the N dSph galaxy  provides
another datum at the low luminosity end, where there is some real scatter,
as well as disagreement about abundances for certain galaxies (e.g. 
\citet{2002AJ....124..886D} find [Fe/H]=-2.2 for And V, while \citet{2005MNRAS.356..979M}
find -1.5).

\begin{figure}
\plotone{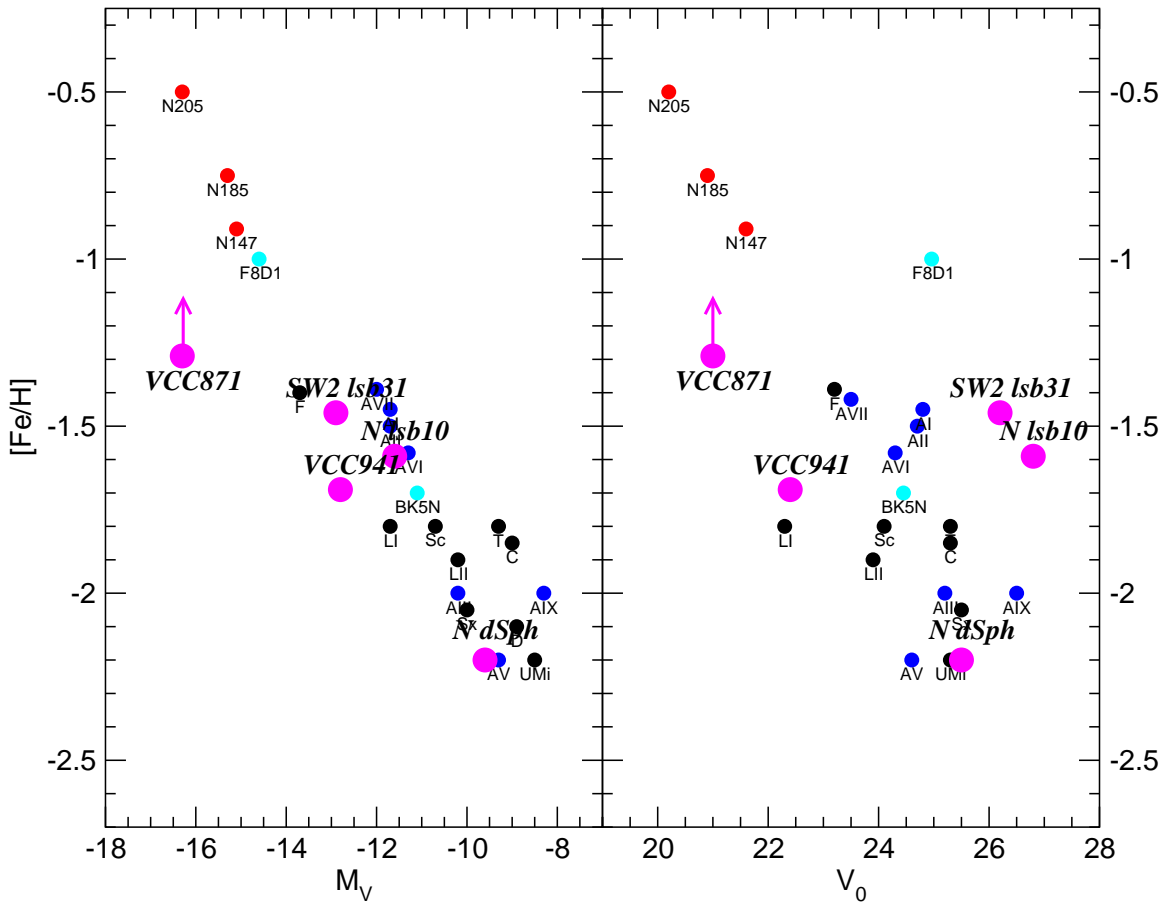}
%\plotone{f11.eps}
\caption{Left: relation of mean [Fe/H] versus absolute V magnitude, for
dwarf ellipticals where the abundance has been determined by the giant branch
color.  The five new Virgo objects are identified in  larger type.  The uncertainties
in [Fe/H] are listed in table \ref{res.tab}, but are not shown here for clarity. 
Right: the 
relation of [Fe/H] versus central V surface brightness. The
[Fe/H] for VCC871 is likely higher than shown here, due to biases discussed in the
text.}
\label{fig:feh_l}
\end{figure}

Figure \ref{fig:feh_l} is a plot showing the relation between
the three parameters of M$_{\rm V}$, V$_{\rm o}$, and [Fe/H]  for
nearby dwarf ellipticals, drawn from the literature.  Plotted as
well, and identified by larger symbols are the 5 dwarf galaxies in 
this study. Clearly, the N lsb10 and SW2 lsb31 dwarfs, the two extremely
low surface brightness galaxies, have abundances typical for their
luminosities, which are brighter than the Sculptor dSph, but fainter
than the Fornax dwarf, even though those two Virgo galaxies have
surface brightnesses nearly 3 magnitudes fainter than those Local Group
galaxies.

The N dSph galaxy, with its luminosity close to that of Draco and Uminor,
appears to have a metal abundance lower than any previously known dwarf,
though of course the uncertainties are high on its abundance.  VCC941, the relatively
high surface brightness galaxy  has an abundance within the spread of more
nearby galaxies of its luminosity, though its metallicity is much lower
than some other galaxies of similar surface brightness.  In that way, it 
is like Leo I, another relatively high surface brightness galaxy whose
abundance tracks luminosity rather than surface brightness.

The photometric bias that resulted in a loss of very red, hence metal rich
stars, shows up particularly in the case of VCC871. Only a 
a lower estimate was found for [Fe/H], and that is from a location
6 kpc or 4.5 $\Re _{\rm eff}$ from the center,  which may be more metal poor than
the galaxy mean.  So the location of the datum for VCC871  serves only to
indicate a lower bound of the true mean abundance.

\section{Summary} 
The resolving power of the ACS on the Hubble telescope allows 
individual red giant
stars in galaxies as distant as the Virgo cluster to be imaged.  Although
only two fields were studied here, and the depth of the F555W images leaves much
to be desired, a number of interesting results have come out.  

Regarding dwarf galaxies, we can say for certain that extremely
low surface brightness galaxies do exist, and can inhabit the central
regions of clusters.   That such apparently tenuous galaxies are found
in densely packed volumes of space is somewhat of a curiosity, considering
the possibilities for disruption due to tidal forces.  The reason for survival
may have to do with the dominance of dark matter in dwarfs, or the orbits of
the dwarfs within the cluster.  Orbits which confine the dwarfs to the cluster core
would result in small tidal stresses due to the mean cluster field, though
tides from interactions with individual galaxies could still be destructive.

The five dwarf ellipticals studied here are similar in luminosity and
metal abundance to
those of the Local Group and other nearby groups, in that metallicity 
closely follows galaxy luminosity, but is nearly independent of surface brightness.
However, the two principal target galaxies, N lsb10 and SW2 lsb31, are examples
of extremely low surface brightness galaxies with limiting radii greater than
2 kpc, and are not currently known in the Local Group.
The lack of a surface brightness-metallicity correlation could mean 
that the large, extremely
low surface brightness galaxies were more compact during their episodes of
star formation, and expanded greatly as a result of either mass loss during those
episodes or external tidal forces.  Why some dwarf galaxies expanded much more than others
is an unresolved question.

The tip of the red giant branch is currently a favored method of deriving distances
for old stellar populations in more nearby galaxies, and thus the resolved
galaxies here are particularly valuable in that they can give a distance to the
Virgo cluster that is independent of previous measurements.  Doing so, 
a mean distance of 16.1 $\pm 0.4$ Mpc is found for the six galaxies observed here, in accord
with the distance of 17.4 $\pm 0.7$ Mpc for the Virgo intracluster stars.
The depth of the cluster is not likely to be more than 1 Mpc, as a gaussian standard
deviation.

Assuming an extended lifetime for HST, other projects to study the stellar populations
in a variety of galaxy types appear possible in Virgo, perhaps providing clues to
the galaxy morphology transformations that occur in cluster environments.

\acknowledgments
Thanks to T.A. Armandroff and G.S. DaCosta for helpful discussions on
dwarf galaxies, Peter Stetson and Lucas Macri on the fine points of
using DAOPHOT on ACS images, and John Roll for the use and abuse of
the Starbase programs.  This project was support under NASA grant
GO-9363 from the Space Telescope Science Institute, which is operated
by AURA, Inc, under NASA contract.

\clearpage
\pagestyle{empty}
%\setcounter{page}{0}
%\vspace*{1.5in}
\begin{deluxetable}{llllllllcccc}
%\rotate
\tablenum{1}
\tablecolumns{11}
\tablewidth{0pc}
\tablecaption{Basic Data and Exposures for Virgo Targets 
\label{data.tab}
}
\tablehead{\colhead{Field} & \colhead{Name} & \colhead{RA}  &\colhead{Dec}
 &\colhead{R}   &\colhead{B--R}    &\colhead{B$_{\rm o}$}   & \colhead{$\Re _{\rm eff}$} & \multicolumn{2}{c} {F555W} & \multicolumn{2}{c}{F814W}  \\
\colhead{} & \colhead{}  &\multicolumn{2}{c}{2000}  & \colhead{}
 &\colhead{}   &\colhead{} &\colhead{arcsec} 
&\colhead{exp (s) }  &\colhead{\# }  &\colhead{exp (s)}   &\colhead{\#} 
}
\startdata
North&N lsb10 & 12:26:48.1 & 13:21:17.5 &$ 18.54\pm.13$& $1.44\pm.04 $& $ 27.64\pm.58 $ & 	19.7  & 2560 &13 &1200 & 12 \\ 
&VCC941 &  12:26:47.9 & 13:22:45.8 &  $17.73\pm.01 $ & $ 1.29\pm.00 $& $23.18\pm.23 $ & 	4.0  &  & &&  \\ 
&N dSph & 12:26:42.1 & 13:22:33.0 & $21.32\pm.05 $& $ 1.51\pm.07 $&  $27.2\pm.3$  & 	5.0\tablenotemark{a}  & & & &  \\
\hline
SW2 &SW2 lsb31 &   12:26:20.1 & 12:34:25.3 & $17.76\pm.05 $& $1.28\pm.03   $& $27.46\pm.16  $  & 	21.0 &2560 &13 &1200 & 12 \\
&VCC871 & 12:26:05.6 &  12:33:35.0 &   $14.25\pm.02 $& $ 1.40\pm.00$& $ 21.84\pm.17 $   & 	19.6 & && &  \\
&NGC4407  & 12:26:32.2 & 12:36:40 & 	\nodata &  	\nodata &  	\nodata   & 	\nodata & & & & \\
\enddata
\tablenotetext{a}{Derived from ACS image}
\end{deluxetable}

%\pagestyle{plaintop}
%\setcounter{page}{0}
%\vspace*{1.5in}
\begin{deluxetable}{lrrrrrrrr}

\tablenum{2}
\tablecolumns{11}
\tablewidth{0pc}
\tablecaption{Derived Distances and Metallicities for Virgo Targets 
\label{res.tab}
}
\tablehead{\colhead{Name} & \colhead{E(B-V)}  &\colhead{TRGB\tablenotemark{a}}
%&\colhead{(m-M)$_{\rm 0}$}  &\colhead{[Fe/H]}  &\colhead{M$_{\rm V}$}   &\colhead{M$_{\rm V,acs}$} &\colhead{V$_{\rm o}$} &\colhead{V$_{\rm o,acs}$}
%}
 &\colhead{(m-M)$_{\rm 0}$}  &\colhead{[Fe/H]}  &\colhead{M$_{\rm V}$\tablenotemark{b}}   &\colhead{M$_{\rm V,acs}$\tablenotemark{c}} &\colhead{V$_{\rm o}$\tablenotemark{b}} &\colhead{V$_{\rm o,acs}$\tablenotemark{c}}
}
\startdata

N lsb10&  	0.03&	27.0& 	31.0$\pm	0.12$&	-1.6$\pm	0.2$&	-11.6&   	\nodata &	26.8&    	26.2\\     
VCC941&   	0.03&	27.1& 	31.1$\pm	0.11$&	-1.7$\pm	0.2$&	-12.9&   	-12.8&   	22.4&    	22.4\\     
N dsph&   	0.03&	27.0& 	30.9$\pm	0.17$&	-2.2$\pm	0.4$&	-8.9&    	-9.4&    	26.4&    	25.5\\     
N back&   	0.03&	27.0& 	31.1$\pm	0.12$&	-1.1$\pm	0.2$&	\nodata &	\nodata &	\nodata &	29.0\\     
SW2 lsb31&	0.03&	27.0& 	30.9$\pm	0.11$&	-1.5$\pm	0.2$&	-12.9&   	\nodata &	26.1&    	26.2\\     
NGC4407&  	0.03&	27.1& 	31.1$\pm	0.12$&	-1.0$\pm	0.3$&	\nodata &	\nodata &	\nodata &	\nodata  \\
VCC871&   	0.03&	27.3& 	31.2$\pm	0.17$&	-1.3$\pm	0.4$&	-16.4&   	\nodata &	21.0&    	\nodata  \\
SW2 back2&	0.03&	27.2& 	31.2$\pm	0.11$&	-1.2$\pm	0.2$&	\nodata &	\nodata &	\nodata &	\nodata  \\
SW2 back4&	0.03&	27.3& 	31.3$\pm	0.11$&	-1.1$\pm	0.1$&	\nodata &	\nodata &	\nodata &	27.5\\     

\enddata
\tablenotetext{a}{Observed I magnitude}
\tablenotetext{b}{derived from B$_{\rm o}$  and extinction from table \ref{data.tab}, and assuming B-V=0.7}
\tablenotetext{c}{Derived from ACS images, either via surface photometry, or 
star counts calibrated to surface photometry (see text).}
\end{deluxetable}

\clearpage



\clearpage

%\vfill\eject

%\clearpage

%\clearpage

\end{document}